\documentclass[final,5p,times]{elsarticle}

\usepackage{hyperref}
\hypersetup{colorlinks=true}
\usepackage{amssymb}
\usepackage{amsmath}
\usepackage{amsthm}
\usepackage{graphicx}
\usepackage{caption}
\usepackage{subcaption}

\journal{Fusion Engineering and Design}

\begin{document}

\begin{frontmatter}

%% Title, authors and addresses

%% use the tnoteref command within \title for footnotes;
%% use the tnotetext command for the associated footnote;
%% use the fnref command within \author or \address for footnotes;
%% use the fntext command for the associated footnote;
%% use the corref command within \author for corresponding author footnotes;
%% use the cortext command for the associated footnote;
%% use the ead command for the email address,
%% and the form \ead[url] for the home page:
%%
%% \title{Title\tnoteref{label1}}
%% \tnotetext[label1]{}
%% \author{Name\corref{cor1}\fnref{label2}}
%% \ead{email address}
%% \ead[url]{home page}
%% \fntext[label2]{}
%% \cortext[cor1]{}
%% \address{Address\fnref{label3}}
%% \fntext[label3]{}

\title{The effect of a micro bubble dispersed gas phase on hydrogen isotope transport in liquid metals under nuclear irradiation.}

%% use optional labels to link authors explicitly to addresses:
%% \author[label1,label2]{<author name>}
%% \address[label1]{<address>}
%% \address[label2]{<address>}

\author[1]{J. Fradera\corref{cor1}\fnref{labelc1}}
\author[1]{S.Cuesta-L\'{o}pez\fnref{labelc2}}
\address[1]{Advanced Materials, Nuclear Technology, Applied Nanotechnology, University of Burgos (UBU), Science and Technology Park, I+D+I Building, Room 63, Plaza Misael Bañuelos, s/n, 09001, Burgos, Spain}
\cortext[cor1]{corresponding author}
\fntext[labelc1]{contact: jfradera@ubu.es}
\fntext[labelc2]{correspondence may also be sent to: scuesta@ubu.es}
\newcommand{\OpF}{OpenFOAM\textsuperscript{\textregistered} }

\begin{abstract}

The present work intend to be a first step towards the understanding and quantification of the hydrogen isotope complex phenomena in liquid metals for nuclear technology. Liquid metals under nuclear irradiation in ,e.g., breeding blankets of a nuclear fusion reactor would generate tritium which is to be extracted and recirculated as fuel. At the same time that tritium is bred, helium is also generated and may precipitate in the form of nano bubbles. Other liquid metal systems of a nuclear reactor involve hydrogen isotope absorption processes, e.g., tritium extraction system. Hence, hydrogen isotope  absorption into gas bubbles modelling and control may have a capital importance regarding design, operation and safety.

Here general models for hydrogen isotopes transport in liquid metal and absorption into gas phase, that do not depend on the mass transfer limiting regime, are exposed and implemented in \OpF CFD tool for 0D to 3D simulations. Results for a 0D case show the impact of a He dispersed phase of nano bubbles on hydrogen isotopes inventory at different temperatures as well as the inventory evolution during a He nucleation event. In addition, 1D and 2D axisymmetric cases are exposed showing the effect of a He dispersed gas phase on hydrogen isotope permeation through a lithium lead eutectic alloy and the effect of vortical structures on hydrogen isotope transport at a backward facing step.

Exposed results give a valuable insight on current nuclear technology regarding the importance of controlling hydrogen isotope transport and its interactions with nucleation event through gas absorption processes.

\end{abstract}

\begin{keyword}
%% keywords here, in the form: keyword \sep keyword
hydrogen isotopes \sep permeation \sep two-phase flow \sep nucleation \sep CFD
%% MSC codes here, in the form: \MSC code \sep code
%% or \MSC[2008] code \sep code (2000 is the default)

\end{keyword}

\end{frontmatter}

\newcommand{\OpF}{OpenFOAM\textsuperscript{\textregistered} }

%%
%% Start line numbering here if you want
%%
% \linenumbers

%% main text
\section*{Glossary}
\newcommand{\Item}[2]{\item[\textbf{#1\hfill}] #2}
\newcommand{\Itema}[2]{\item[\hfill #1] #2}

\subsection*{Abbreviations}
\begin{list}{}{%
\settowidth{\labelwidth}{\textbf{$A,B,C,D$}}%
\setlength{\labelsep}{2. em}%
\setlength{\leftmargin}{\labelwidth}%
\addtolength{\leftmargin}{\labelsep}%
\setlength{\rightmargin}{0. cm}%
\setlength{\parsep}{\parskip}%
\setlength{\itemsep}{0. cm}\setlength{\topsep}{0. cm}%
\setlength{\partopsep}{0. cm}}

\Item{BFS}{Backward Facing Step}
\Item{CFD}{Computational Fluid Dynamics}
\Item{EoS}{Equation of State}
\Item{HCLL}{Helium Cooled Lithium Lead}
\Item{LM}{Liquid Metal}
\end{list}

\subsection*{Greek characters}
\begin{list}{}{%
\settowidth{\labelwidth}{\textbf{$A,B,C,D$}}%
\setlength{\labelsep}{2. em}%
\setlength{\leftmargin}{\labelwidth}%
\addtolength{\leftmargin}{\labelsep}%
\setlength{\rightmargin}{0. cm}%
\setlength{\parsep}{\parskip}%
\setlength{\itemsep}{0. cm}\setlength{\topsep}{0. cm}%
\setlength{\partopsep}{0. cm}}

\Item{$\alpha$}{void fraction}
\Item{$\delta$}{diffusion layer thickness}
\Item{$\pi$}{number pi}

\end{list}

\subsection*{Latin characters}
\begin{list}{}{%
\settowidth{\labelwidth}{\textbf{$A,B,C,D$}}%
\setlength{\labelsep}{2. em}%
\setlength{\leftmargin}{\labelwidth}%
\addtolength{\leftmargin}{\labelsep}%
\setlength{\rightmargin}{0. cm}%
\setlength{\parsep}{\parskip}%
\setlength{\itemsep}{0. cm}\setlength{\topsep}{0. cm}%
\setlength{\partopsep}{0. cm}}

\Item{$a$}{specific area}
\Item{$k_{d}$}{dissociation coefficient}
\Item{$k_{r}$}{recombination coefficient}
\Item{$k_{S}$}{Sievert's coefficient}
\Item{$p$}{partial pressure}
\Item{$r$}{radial coordinate}
\Item{$r_b$}{bubble radius}
\Item{$t$}{time}
\Item{$\textbf{u}$}{velocity}
\Item{$C$}{concentration}
\Item{$D$}{diffusivity}
\Item{$J$}{molar flux}
\Item{$M$}{molar mass}
\Item{$P$}{pressure}
\Item{$R$}{gas constant}
\Item{$S$}{source term}
\Item{$T$}{temperature}
\Item{$\hat{V}$}{molar volume}
\Item{$Z$}{compressibility factor}

\end{list}

\subsection*{Subscripts}
\begin{list}{}{%
\settowidth{\labelwidth}{\textbf{$A,B,C,D$}}%
\setlength{\labelsep}{2. em}%
\setlength{\leftmargin}{\labelwidth}%
\addtolength{\leftmargin}{\labelsep}%
\setlength{\rightmargin}{0. cm}%
\setlength{\parsep}{\parskip}%
\setlength{\itemsep}{0. cm}\setlength{\topsep}{0. cm}%
\setlength{\partopsep}{0. cm}}

\Item{$abs$}{absorption into helium bubbles}
\Item{$G$}{gas phase}
\Item{$He$}{helium}
\Item{$i$}{hydrogen isotope}
\Item{$LM$}{Liquid Metal}
\Item{$LBE$}{Lead Bismuth Eutectic alloy}
\Item{$LLE$}{Lithium Lead Eutectic alloy}

\end{list}

\section{\label{sec:intro}Introduction}

Despite the technological interest of the interaction of hydrogen with matter in different fields (i.e. study of hydrogen embrittlement in many industrial applications such as petrochemical plants, chemical reactors, pipe lines ), hydrogen isotope transport in matter is a critical question in current nuclear technologies, from the point of view of design, operation and safety issues.
In thermonuclear fusion devices, the fuel is a high temperature deuterium-€"tritium plasma. It is important to predict and control the deuterium and tritium Inventory in, Permeation through and Recycling from (IPR) the reactor walls, where, in general, inventory and permeation should be minimized. Moreover, understanding, controlling and therefore predicting H transport phenomena is a key requisite in the design of core fusion reactor components like He Cooled Lithium Lead (HCLL) Breeding Blankets (BB) (see, e.g., \citet{Salavy} for details on HCLL design), extraction systems of future nuclear fusion reactors, high energy conversion fission reactor systems involving Lead Bismuth Eutectic (LBE) alloy and steam, and liquid metal experimental facilities.

One of the current unsolved questions is the influence that a gas phase in the form of bubbles, either nucleated in the bulk material or injected for solute extraction purposes, may have on hydrogen isotopes inventory and transport parameters (see \citet{Norajitra}).

The availability of a computational tool fully devoted to hydrogen isotopes transport evaluation in nuclear materials, specially in liquid metals, is of great importance, e.g., tritium inventory control and confinement is a key issue in nuclear fusion D--T reactors, concerning safety and the fuel cycle. Well-known transport models have been adapted and implemented for CFD, taking into account interface mass transfer for hydrogen isotopes, gas bubble nucleation and growth, and transport of gas bubbles.

\begin{figure}
\begin{center}
\includegraphics[angle=0,width=0.5\columnwidth]{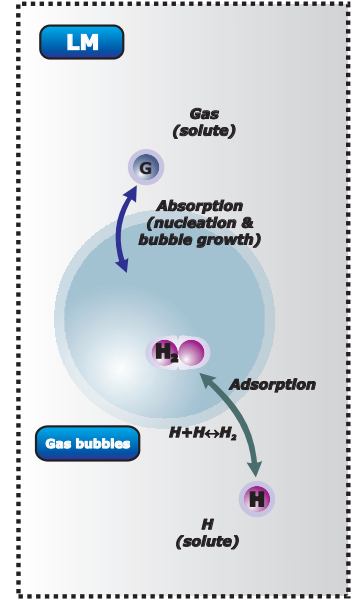}
\caption{Hydrogen isotopes transport phenomena in the presence of a gas phase in the form of bubbles.}
\label{fig:Phenomena}
\end{center}
\end{figure}

Present work intends to be a step forward towards the quantification of the complex phenomena of hydrogen isotopes transport in liquid metals (as fusion materials) focusing on a gas phase within the liquid metal bulk phase in the form of micro bubbles Fig.~\ref{fig:Phenomena} shows a representation of the hydrogen isotopes transport phenomena in the presence of a gas phase in the form of bubbles. Gas bubble species (e.g., helium), in the form of a solute in a liquid metal, can be absorbed and desorbed from the bubbles making these grow or shrink. In addition, hydrogen isotopes are adsorbed into the gas bubbles.

In this paper, we introduce a new model aiming to describe such complex phenomena, improving previous work \citep{Fradera11} where a simple and specific model was presented for tritium transport in HCLL BBs with nucleation. In the present work new considerations have been implemented and adapted to CFD in order to ensure more reliable and predictive simulations.

The model described in Section 2 has taken into account the following phenomena:

\begin{itemize}
\setlength{\itemsep}{1pt}
\setlength{\parskip}{0pt}
\setlength{\parsep}{0pt}
 \item{Hydrogen isotopes diffusion and convection in liquid metals. Sec.~\ref{sec:Diffusion}.}\label{Ph_1}
 \item{Hydrogen isotopes absorption into gas bubbles. Sec.~\ref{sec:HAbsorption}.}\label{Ph_2}
 \item{Helium nucleation diffusion and convection in liquid metals. Sec.~\ref{sec:Helium}.}\label{Ph_3}
 \item{Helium absorption: bubble growth. Sec.~\ref{sec:HeAbsorption}.}\label{Ph_4}\\
\end{itemize}

Hydrogen isotopes transport phenomena in the presence of gas bubbles are analysed by means of a novel solver for \OpF CFD open source code \citep{Jasak}.

In addition, in Section 4, we discuss the following cases are exposed for analysis and raise conclusions related to current fusion technology designs and operation conditions:

\begin{enumerate}
\setlength{\itemsep}{1pt}
\setlength{\parskip}{0pt}
\setlength{\parsep}{0pt}
 \item{Hydrogen isotopes diffusion in Lithium Lead eutectic (LLE) alloy with a homogeneous dispersed phase consisting in micro-bubbles Sec.~\ref{sec:case1}.}\label{0D_1case}
 \item{Irradiation induced nucleation event in LLE Sec.~\ref{sec:case2}.}\label{0D_2case}
 \item{Hydrogen isotopes permeation through LLE with a homogeneous dispersed phase consisting in micro-bubbles Sec.~\ref{sec:case3}.}\label{1D_1case}
 \item{Hydrogen isotopes diffusion in LLE flowing through a sudden expansion Sec.~\ref{sec:case4}.}\label{2D_1case}\\
\end{enumerate}

Sec.~\ref{sec:case1} case presents the hydrogen transport phenomena in the presence of a homogeneous concentration of He micro-bubbles. The absence of flow or concentration gradients shows the hydrogen absorption effect alone. Therefore, pure effect of absorption in hydrogen isotopes inventory can be analysed.

LLE would be used as a tritium breeding material in future fusion reactors. Helium and tritium are generated in the bulk LLE at the same rate by nuclear reactions from $^6$Li. Sec.~\ref{sec:case2} case shows the transient effect on hydrogen absorption while a nucleation event due to He accumulation takes place.

Sec.~\ref{sec:case3} case exposes the delay effect on hydrogen isotopes diffusion through a LLE 1D slab in the presence of a homogeneous dispersed gas phase in the form of He micro bubbles.

Low viscosity and high density of liquid metals like LLE promote the formation of vortices. A classic sudden expansion, showing the effect of well-known and studied vortices or recirculation zones that this type of geometry generates on hydrogen isotopes transport in the presence of helium micro-bubbles is exposed in Sec.~\ref{sec:case4}.

\section{Implemented model}

An adaptation of classic and well known models to a CFD code have been implemented with the aim of predicting and analyse the effect of a gas phase in the form of micro bubbles on hydrogen isotope transport phenomena. Hydrogen isotope absorption into the gas phase can modelled assuming either Diffusion Limited Regime (DLR), i.e. Sieverts' law, or Surface Limited Regime (SLR). In this work, a series of resistances model has been implemented. Impact of absorbed hydrogen on bubbles volume have been neglected as discussed in ~\ref{App:App1}.

Gas bubble nucleation has been modelled with the Self-Consistent Nucleation Theory (see details in ~\ref{App:App2}).Finally, bubbles are treated as a passive scalar, so, hydrogen isotope absorption is calculated as a source term. All models have been coupled taking into account any interaction between the aforementioned phenomena.

\subsection{Hydrogen diffusion and convection in liquid metals}\label{sec:Diffusion}

Hydrogen isotope transport in liquid metals can be modelled with the following governing equations (microscopic mass balance equations):

\begin{eqnarray}
\dfrac{\partial{C_{i,LM}}}{\partial{t}} &=& -\textbf{u}\cdot\nabla C_{i,LM}+\nabla D_{i,LM} \nabla C_{i,LM} +S_{i,gen}-S\!_{i,abs}
\\
\dfrac{\partial{C_{i_2,G}}}{\partial{t}} &=& -\textbf{u}\cdot\nabla C_{i_2,G}+\dfrac{1}{2}S\!_{i,abs}\label{eq:CT2G}
\end{eqnarray}

where $C_i$ is the hydrogen isotope concentration in atomic form, $C_{i_2}$ is the hydrogen isotope concentration in molecular form (gas phase) and $D$ is the diffusion coefficient, subscript $G$ stands here for the whole gas phase and subscript $LM$ denotes the liquid metal phase. All concentrations are referred to the LM volume. $\textbf{u}$ is the LM velocity. Gas phase is treated as a passive scalar so the gas phase moves along with the LM. $S_{i,gen}$ is a source term taking into account possible hydrogen isotope generation due to nuclear reactions or other sources. $S_{i,abs}$ is a source term taking into account the absorption or desorption processes due to the presence of micro bubbles.

Hydrogen isotope trapping effect, has been taken into account through an effective diffusion coefficient (see, e.g., \citet{Esteban} for more details).It is worth noting that hydrogen isotopes dissolved in a LM behave as non volatile gases, so they do not nucleate unless extreme temperature and pressure conditions are met.

\subsection{Hydrogen isotope absorption into helium gas bubbles}\label{sec:HAbsorption}

\begin{figure}
\begin{center}
\includegraphics[angle=0,width=1\columnwidth]{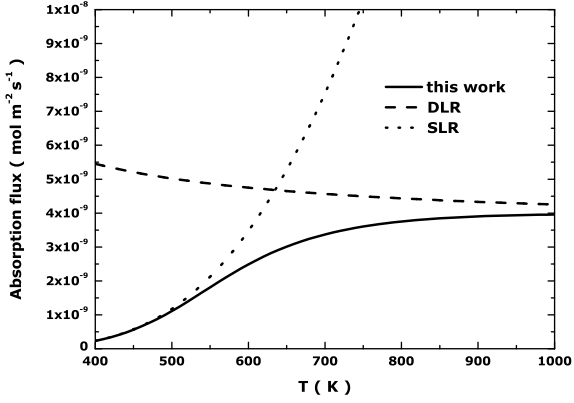}
\caption{Absorption flux predicted as a function of temperature comparing the model analysed in this work to models assuming DLR and SLR conditions. Parameters $D_{H,LM}$ and $k_s$ from \citep{Reiter}, $k_r$ and $k_d$ from \citep{Pisarev},  $C_{H,LM} =$ 10$^{-4}$ mol/m$^3$, $p_{H_2} =$ 10 Pa}
\label{fig:regime}
\end{center}
\end{figure}

Absorption of diatomic gases like hydrogen isotopes in LM, can be modelled as series of resistances: diffusion through a stagnant layer around the gas phase and surface recombination--dissociation process at the LM-helium gas phase interface. Thus, process can follow a DLR o SLR depending on the gas phase pressure and surface conditions or a combination of the aforesaid resistances can be controlling the phenomenon (see Fig.~\ref{fig:regime}).

Assuming SLR, hydrogen isotopes flux through the interface (mol/m$^2$s) can be expressed as follows:
\begin{equation}\label{eq:SLRflux}
J_{i,LM \rightarrow G}=-k_{r}C_{i,LM \rightarrow G}^{2}+k_{d}^{2}C_{i_2,G \rightarrow LM}
\end{equation}
where $k_{r}$ is the recombination coefficient and $k_{d}$ is the dissociation coefficient, both at the LM-gas interface. $G$ stands for the He gas phase.

If DLR is assumed, Sievert's law applies at the LM-gas interface, i.e., the interface is always in equilibrium and mass transfer if controlled by the diffusion in a stagnant layer around the interface; surface process resistance is negligible.
\begin{equation}\label{eq:Sieverts}
C_{i,LM \rightarrow G}=k_{s}p_{i_2}^{1/2}
\end{equation}
where $p_{i_2}$ is the partial pressure of molecular $i_2$ inside the gas phase and $k_{s}$ is the  Sieverts' coefficient for $i$ in the specific LM.

Hence, the flux (mol/m$^2$s) of H through the LM-gas phase interface can be expressed as follows:
\begin{equation}\label{eq:DLRflux}
J_{i,LM \rightarrow G}=\dfrac{D_{i,LM}}{\delta}(C_{i,LM \rightarrow G}-k_{s}p_{i_2}^{1/2})
\end{equation}
where $\delta$ is the thickness of the LM diffusion layer around the interface.

Provided that realistic system conditions may not be either DLR or SLR, a flux, taking into account both processes (see ~\ref{App:App3}), reads,
\begin{eqnarray}\label{eq:flux}
&J_{i,LM \rightarrow G}=\dfrac{D_{i,LM}}{\delta}\Bigg[C_{i,LM \rightarrow G}-\nonumber\\
&\Bigg(\dfrac{-D_{i,LM}}{2k_r\delta}+\sqrt{\left(\dfrac{D_{i,LM}}{2k_r\delta}\right)^2+\dfrac{-D_{i,LM}C_{i,LM \rightarrow G}}{k_r\delta}+\dfrac{k_d p_{i_2}}{k_r}}\Bigg)\Bigg]\nonumber\\
\end{eqnarray}

As Fig.~\ref{fig:regime} shows, this flux model (Eq.~\ref{eq:flux}) represents a consistent description covering intermediate conditions at time of representing system conditions that follow DLR (Eq.~\ref{eq:DLRflux}), or SLR(Eq.~\ref{eq:SLRflux}) regimes.

Hence, the rate of hydrogen isotope absorption reads,
\begin{equation}
S_{i,abs} = J_{i,LM \rightarrow G} A_G\label{eq:SHAbs}
\end{equation}
where $A_G$ is the total interface area of the gas phase.

\subsection{Helium diffusion and convection in liquid metals}\label{sec:Helium}

The He transport model proposed is based on the assumption that bubbles are very small (fluid density is that of pure liquid) and they move along with the fluid, so, in the present work, a passive scalar modelling approach has been implemented. This approach assumes that bubbles movement has no effect on the flow field, which is reasonable due to the size of the bubbles and the low viscosity and high density of the fluid. 

Governing equations are formulated for dissolved atomic He ($C_{He,LM}$), for He in the gas phase ($C_{He,G}$) and for the number of bubbles per unit volume ($N_b$):
\begin{equation}
\dfrac{\partial{C_{He,LM}}}{\partial{t}}\ =\ -\textbf{u}\cdot\nabla C_{He,LM} + \nabla D_{He,LM}\nabla C_{He,LM}+ S_{He} - S_{He,abs}-S_{nuc}\label{eq:CHeLM}
\end{equation}
\begin{equation}
\dfrac{\partial{C_{He,G}}}{\partial{t}}\ =\ -\textbf{u}\cdot\nabla C_{He,G} + S_{He,abs}+ S_{nuc}\label{eq:CHeG}
\end{equation}
\begin{equation}
\dfrac{\partial{N_{b}}}{\partial{t}}\ =\ -\textbf{u}\cdot\nabla N_{b}+S_{nuc}\label{eq:Nb}
\end{equation}
where $S_{He,gen}$ is a source term taking into account He generation, e.g., by nuclear reactions, $S_{nuc}$ is a source term taking into account nucleation and $S_{He,abs}$ is the rate of He absorption due to the bubble growth mechanism. Note that all concentrations are referred to the LM volume so, e.g., $C_{He,G}$ is the He concentration in the gas phase per LM volume.

\subsection{Helium absorption into gas bubbles}\label{sec:HeAbsorption}

Absorption of noble gases like Helium can be modelled as a DLR as there is no recombination-dissociation process at the LM-gas phase interface. Hence, assuming that inertial effects can be neglected due to the size of the bubbles (micro bubbles), the He flux through the interface of one bubble $J_{He,LM \rightarrow b}$ can be calculated as follows:
\begin{equation}
J_{He,LM \rightarrow b} = D_{He,LM}\,\biggl(\dfrac{\partial C_{He,LM}}{\partial r}\biggl)_{r=r_{b}}\label{eq:JHeAbs}
\end{equation}
where $r_b$ is the radius of a bubble.

The concentration gradient $\left( \partial C_{He,LM}/\partial r \right)_{r=r_{b}}$ is approximated to: 
\begin{equation}
\biggl(\dfrac{\partial C_{He,LM}}{\partial r}\biggl)_{r=r_{b}}\approx \frac{C_{He,LM}-C^{sat}_{He,LM}}{r_{b}}\label{eq:Cgrad}
\end{equation}
where $C^{sat}_{He,LM}$ is the He saturation concentration. Note that Eq.~\ref{eq:Cgrad} is a simplification of the \citet{Epstein} model.

Hence, the rate of He absorption reads,
\begin{equation}
S_{He,abs} = J_{He,LM \rightarrow b} N_b A_b\label{eq:SHeAbs}
\end{equation}
where $A_b$ is the interface area of a bubble.

\section{Models implementation in \OpF}

\OpF (Open Field Operation and Manipulation) is an open source multiphysics CFD code (\citep{OpenFOAM}, \citep{Jasak}), which uses the Finite Volume Method (FVM). The code is produced by OpenCFD Ltd and is freely available and open source, licensed under the GNU General Public Licence. As an open source code, \OpF is a good alternative to commercial CFD codes like Fluent\textsuperscript{\textregistered} \citep{FLUENT} or CFX\textsuperscript{\textregistered} \citep{CFX}. One advantage over the commercial codes is that the source code can be modified, so models can be coded in a rather simple way. \OpF is a flexible and easy to upgrade code, consisting in a set of efficient C++ modules, that let the user add as many physics as needed, i.e. transport equations or EoS.

In the exposed model, the absorption process is treated as a zero dimensional process in every LM control volume, taking the form of a source term (eqs.~\ref{eq:SHAbs} or~\ref{eq:SHeAbs}). Diffusion layer around the interface, $\delta$, can be approximated as $\delta = r_b$ (bubble radius), following \citep{Epstein} approach and neglecting the transient terms in the rate of absorption. Bubbles are not simulated as a separated gas phase but as a passive scalar as it is assumed that their size is so small that they do not affect the LM.

Source terms for T and T$_2$ calculation may introduce numerical instabilities when solving the corresponding governing equations (see Sec.~\ref{sec:Diffusion}). The fractional step method (see \citep{Ferziger} for more detail on this method), first introduced independently by \citep{Chorin} and \citep{Temam}, is used to avoid such instabilities and to have a more flexible and scalable code. A two step fractional step method is used for each transported variable, first solving the source terms (eq.~\ref{eq:CTLMstep1} and eq.~\ref{eq:CT2Gstep1}) and afterwards the convection, diffusion and generation terms (eq.~\ref{eq:CTLMstep2} and eq.~\ref{eq:CT2Gstep2}) of the governing equations as follows:
\begin{eqnarray}
\dfrac{\partial{C_{i,LM}}}{\partial{t}} &=& -S\!_{i,abs}\label{eq:CTLMstep1}\\
\dfrac{\partial{C_{i_2,G}}}{\partial{t}} &=& \dfrac{1}{2}S\!_{i,abs}\label{eq:CT2Gstep1}\\
\nonumber\\
\dfrac{\partial{C_{i,LM}}}{\partial{t}} &=& -(\textbf{u}\cdot\nabla C_{i,LM})+(\nabla D_{i,LM}\nabla C_{i,LM})+S_{i,gen}\label{eq:CTLMstep2}\\
\dfrac{\partial{C_{i_2,G}}}{\partial{t}} &=& -(\textbf{u}\cdot\nabla C_{i_2,G})\label{eq:CT2Gstep2}
\end{eqnarray}

Note that for $C_{i_2,G}$ neither diffusion nor generation terms exist, as bubbles are treated as a passive scalar and $i_2$ is not generated, e.g., by nuclear reactions, in the gas phase.

\section{Analysis and Discussion}

We present a set of representative cases of hydrogen isotope transport phenomena, to analyse and predict the effect of the presence of micro bubbles in the hydrogen isotopes transport coefficients, in conditions of relevance in nuclear technology. All cases have been run with \OpF CFD code developed solver.

\subsection{Zero dimensional analysis of hydrogen isotope absorption into a micro bubble dispersed gas phase}\label{sec:case1}

In this numerical experiment, designed to show the impact of a micro bubble dispersed phase on hydrogen isotopes transport parameters, a 0D LLE domain, with no LM velocity (no convection) is simulated. Hydrogen isotopes concentration is set to an initial value of 10$^{-5}$bubbles/(m$^3$). The number of bubbles is set to a constant value of 10$^{-5}$bubbles/(m$^3$) and the He concentration is set to the saturation concentration so as to avoid bubble growth or dissolution. Pressure is set to constant values of 2~bar. Note that simulation conditions have been chosen to be representative of those of a HCLL BB at operation conditions (see \citep{Salavy} and references therein for more details).

Diffusion and Sievert's coefficients for hydrogen isotopes in LLE have been taken from \citet{Reiter}. Hydrogen absorption parameters in He have been taken from \citep{Terai90, Terai91} and \citep{Pisarev}, who modelled T release from molten LLE and compared results with \citet{Terai91} experimental data with good agreement. It must be noted that there is abundant literature on hydrogen isotopes transport parameters. However, transport coefficients show a wide span of values, specially for the solubility coefficient.

\begin{figure*}
\begin{center}
	\begin{subfigure}[b]{0.475\textwidth}
		\includegraphics[angle=0,width=\textwidth]{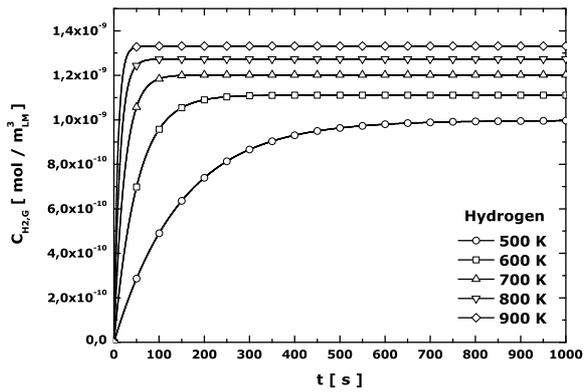}	
		\caption{}	
		\label{fig:hydrogen_T}
	\end{subfigure}	
	\begin{subfigure}[b]{0.475\textwidth}
		\includegraphics[angle=0,width=\textwidth]{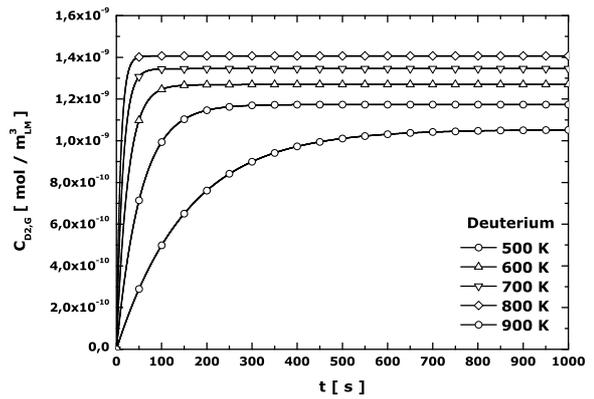}
		\caption{}
		\label{fig:deuterium_T}
	\end{subfigure}
	
	\begin{subfigure}[b]{0.475\textwidth}
		\includegraphics[angle=0,width=\textwidth]{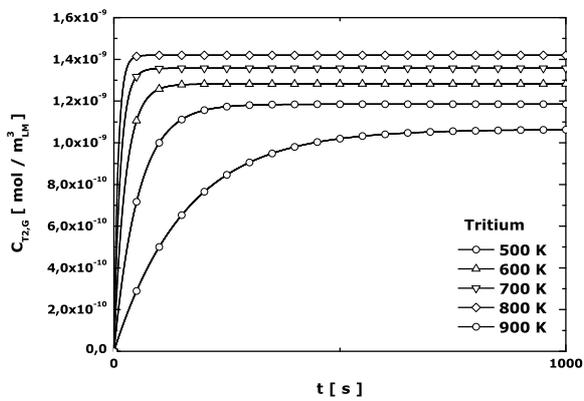}
		\caption{}
		\label{fig:tritium_T}
	\end{subfigure}
	\begin{subfigure}[b]{0.475\textwidth}
		\includegraphics[angle=0,width=\textwidth]{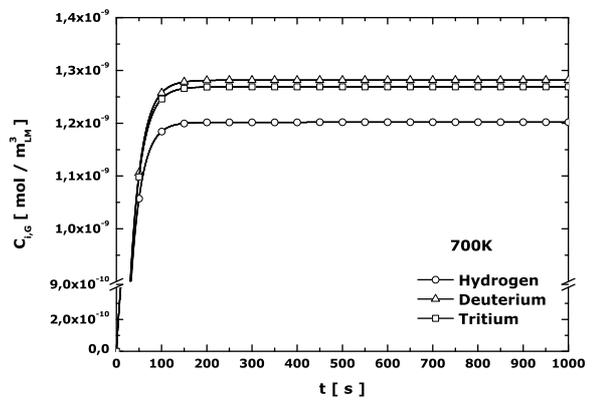}
		\caption{}
		\label{fig:comparison}
	\end{subfigure}
	\caption{(a) Sensitivity to system temperature for the Hydrogen concentration in the gas phase.(b) Sensitivity to system temperature for the Deuterium concentration in the gas phase.(c) Sensitivity to system temperature for the Tritium concentration in the gas phase.(d) Comparison of concentrations in the gas phase for the different hydrogen isotopes at T=700K.}
\end{center}
\end{figure*}

Our results confirm that absorption process increases with increasing temperature. Therefore, bubbles reach hydrogen isotope saturation concentration faster as it is shown in figs.~\ref{fig:hydrogen_T},~\ref{fig:deuterium_T} and fig.~\ref{fig:tritium_T}. Isotope concentration inside the gas phase can be assumed to be the saturation concentration for most engineering applications. However, for fast and unsteady processes, specially at low temperatures, transient absorption may be taken into account.

Fig.~\ref{fig:comparison} shows the gas phase concentration comparison between different isotopes at a given temperature of 700K. Saturation concentration is reached after 200s for all the isotopes, showing no significant difference in the transient absorption process. Different saturation concentrations are reached for each isotope, but the difference may be negligible for transient systems from the inventory account point of view.

\begin{figure*}
\begin{center}
	\begin{subfigure}[b]{0.475\textwidth}
			\includegraphics[angle=0,width=\textwidth]{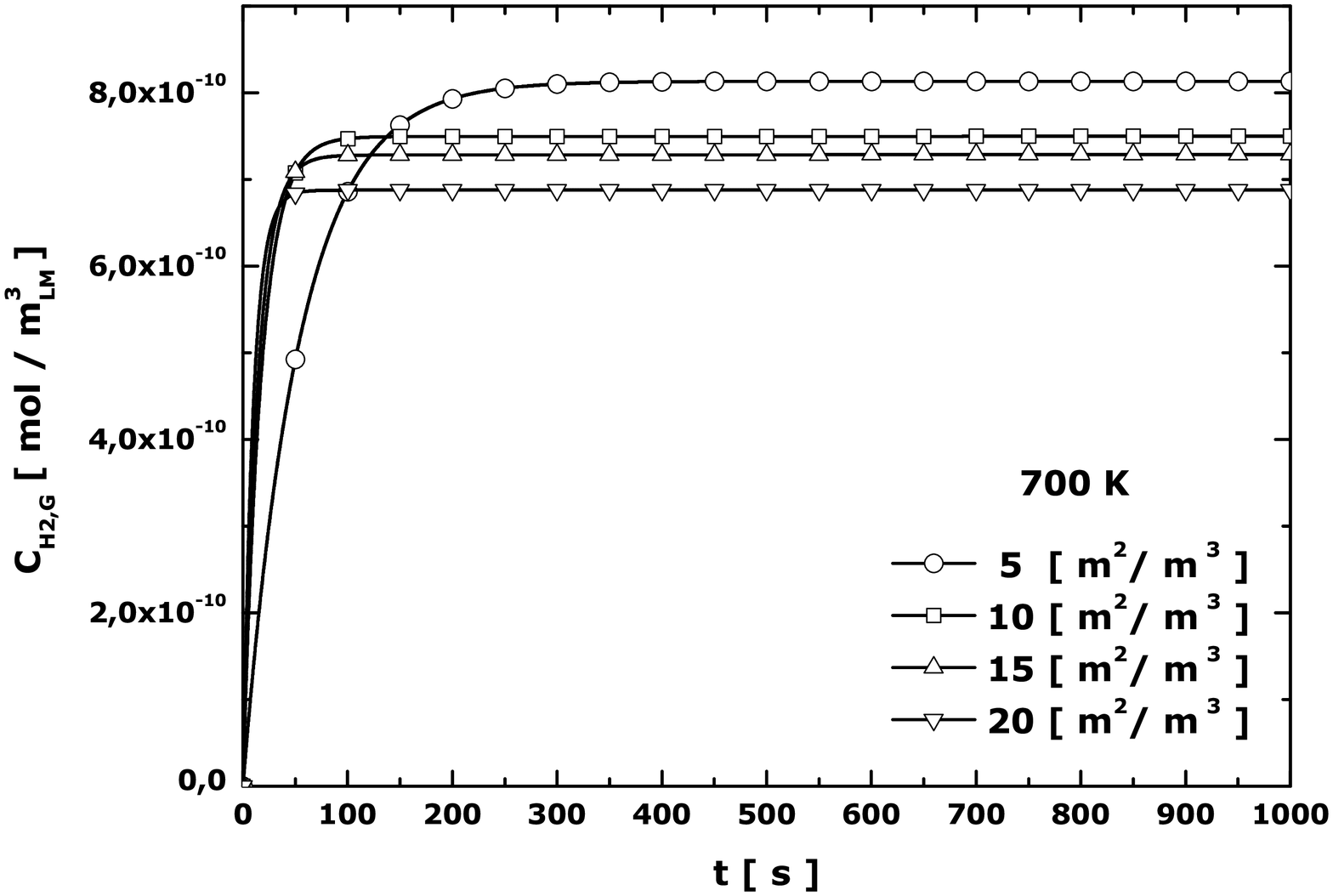}
			\caption{}
			\label{fig:hydrogen_A}
	\end{subfigure}		
	\begin{subfigure}[b]{0.475\textwidth}
			\includegraphics[angle=0,width=\textwidth]{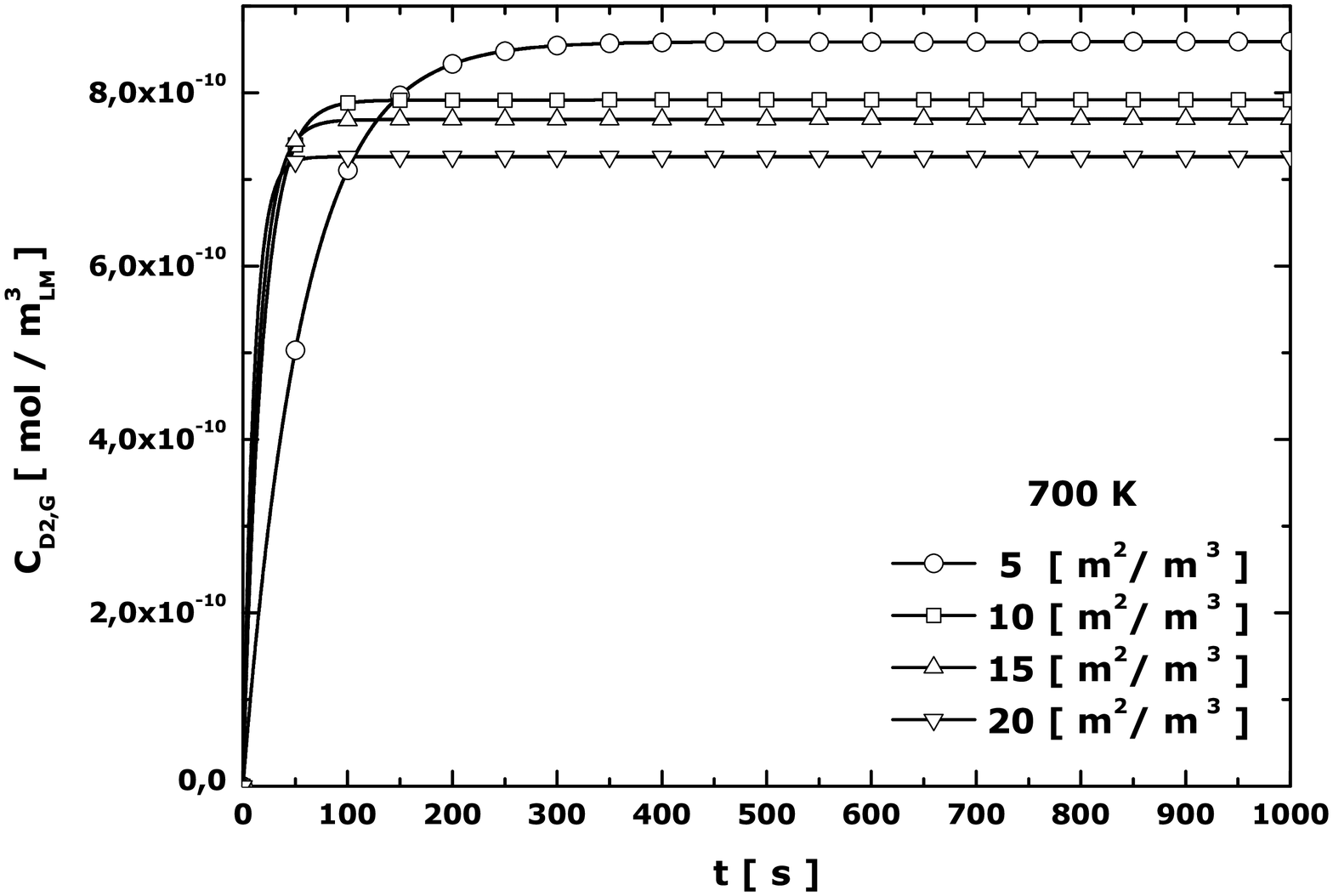}
			\caption{}
			\label{fig:deuterium_A}
	\end{subfigure}
	
	\begin{subfigure}[b]{0.475\textwidth}
			\includegraphics[angle=0,width=\textwidth]{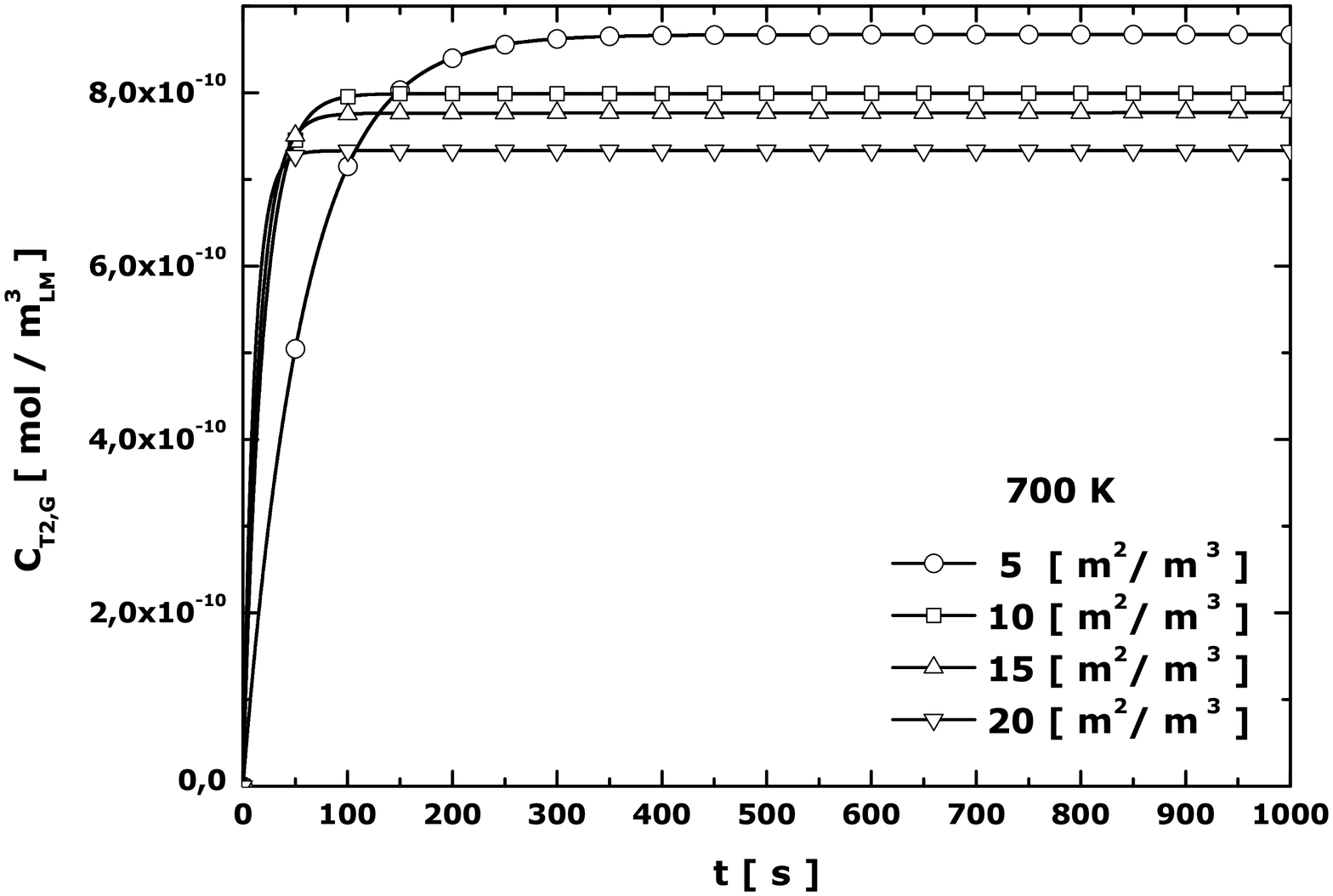}
			\caption{}
			\label{fig:tritium_A}
	\end{subfigure}
	\caption{(a) Hydrogen concentration in the gas phase sensitivity to gas phase specific area. (b) Deuterium concentration in the gas phase sensitivity to gas phase specific area. (c) Tritium concentration in the gas phase sensitivity to gas phase specific area.}
\end{center}
\end{figure*}

As for different bubble concentrations, that is different specific areas $a$ m$^2/$m$^3$, the larger the interface is the higher the amount of absorbed hydrogen isotope. Figs.~\ref{fig:hydrogen_A},~\ref{fig:deuterium_A} and fig.~\ref{fig:tritium_A} show the evolution of the hydrogen isotope gas concentration for different specific areas. Note that saturation concentration is reached much faster for larger specific areas. Therefore, for larger bubbles or larger amounts of bubbles, it can be assumed that hydrogen isotope concentration will be in equilibrium with the hydrogen isotope concentration in the liquid metal.

\subsection{Zero dimensional analysis of hydrogen isotope absorption into a nucleating micro bubble dispersed gas phase}\label{sec:case2}

In this case, the 0D LLE domain in Case ~\ref{sec:case1} has not been populated with bubbles, but a He and hydrogen isotope source term (due to nuclear reactions with Li in the LLE) of 10$^{-7}$mol/(m$^3$) is set to eventually promote a nucleation event. The other parameters have been set as for Case ~\ref{sec:case1} except temperature, that has been set to 600K.

\begin{figure}
\begin{center}
\includegraphics[angle=0,width=1\columnwidth]{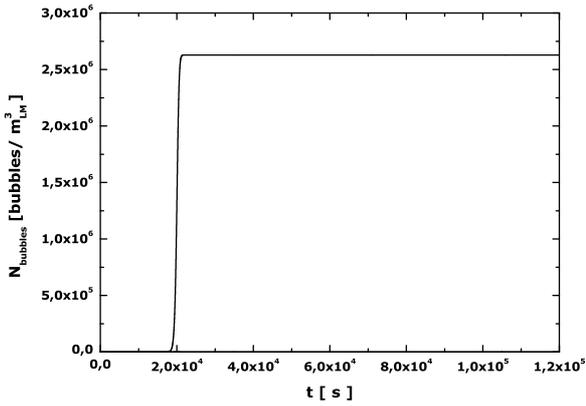}
\caption{Bubble concentration in the liquid metal evolution for the nucleation event at 600K.}
\label{fig:Nb}
\end{center}
\end{figure}

\begin{figure}
\begin{center}
\includegraphics[angle=0,width=1\columnwidth]{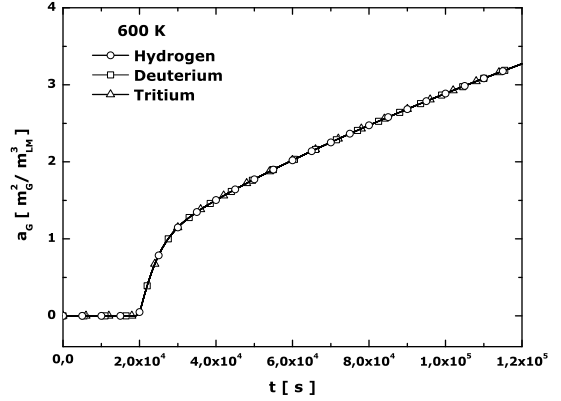}
\caption{Gas phase specific area case comparison for the different absorbed hydrogen isotopes at 600K.}
\label{fig:a}
\end{center}
\end{figure}

We observed that He concentration accumulates until supersaturation is reached. Once the supersaturation reaches the nucleation onset point, that is the nucleation barrier is overcome, nucleation begins. Bubble concentration and gas phase specific area evolution are shown in Fig.~\ref{fig:Nb} and Fig.~\ref{fig:a}, respectively. Specific area grows fast as the nucleation event goes on due to new bubbles, but after a while, dissolved He is almost depleted from the LLE and nucleation stops. Fig.~\ref{fig:Nb} shows how a constant bubble concentration is reached shortly after the nucleation event begins. Once the nucleation event is over, He bubbles grow due to He absorption at a lower rate (see  Fig.~\ref{fig:a} change in slope around 3x10$^{4}$s).

\begin{figure}
\begin{center}
\includegraphics[angle=0,width=1\columnwidth]{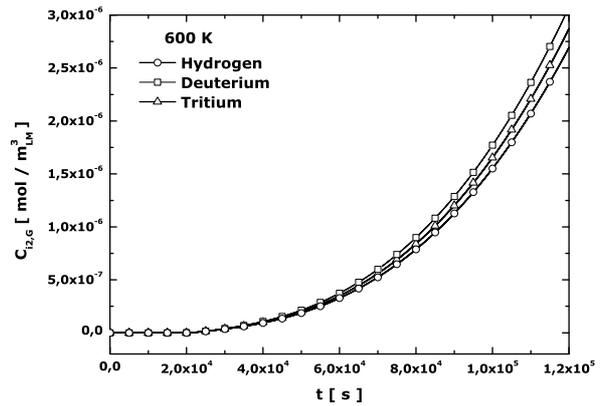}
\caption{Gas phase isotope concentration evolution comparison at 600K.}
\label{fig:CTG0D}
\end{center}
\end{figure}

As bubbles are generated in the LLE, hydrogen isotopes are absorbed into them. In this case, saturation concentration is never reached as bubbles keep growing through He and hydrogen isotopes absorption, as shown in Fig.~\ref{fig:CTG0D}. Note that despite which isotope is present in the LLE, the absorption process does not affect the specific area (see Fig.~\ref{fig:a}) as exposed in ~\ref{App:App1}. However, the isotope concentration in the gas phase is different for each isotope.

Results show that in the presence of a nucleation event or a growing gas phase, hydrogen isotope inventory in the gas phase should be taken into consideration.

\subsection{One dimensional analysis of hydrogen isotope diffusion through LLE in the presence of a micro bubble dispersed gas phase}\label{sec:case3}

Hydrogen permeation through a 1D LLE slab of 1cm length with no convection is set up for this case. Domain is discretized in 10 nodes. The hydrogen isotopes initial concentration in the slab is set to zero. The number of bubbles is set to a constant value of 10$^{7}$bubbles/(m$^3$) (a=20m$^2$/m$^3$) and the He concentration is set to the saturation concentration so as to avoid bubble growth or dissolution. Hydrogen isotope concentration has been set to a constant value of 10$^{-3}$mol/(m$^3$) at x=0m and to zero at x=1cm. The other parameters have been set as for Case ~\ref{sec:case1}.

The well-known unsteady analytical solution \cite{crank} for the flux at x=1cm, that is the flux leaving the system reads, 
\begin{equation}\label{eq:analytical}
J_{i,x=L}(t)=\dfrac{D_iC_{i,x=0}}{L}\left[1+2\sum_{n=1}^{\infty}(-1)^{n}e^{\dfrac{-D_i\pi^2n^2t}{L^2}}\right]
\end{equation}

\begin{figure}
\begin{center}
\includegraphics[angle=0,width=1\columnwidth]{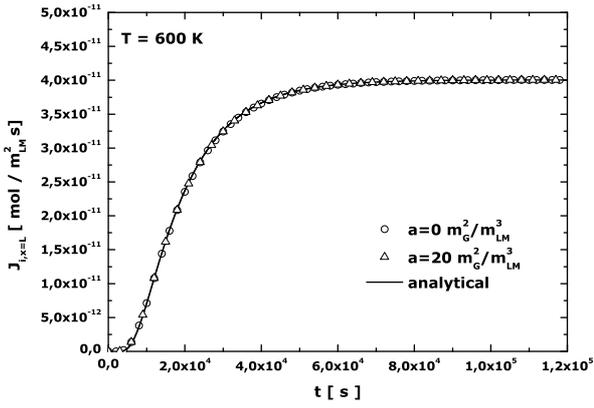}
\caption{Permeation flux evolution comparison between the analytical solution in the absence of bubbles and simulations without bubbles and with a significant specific area of bubbles.}
\label{fig:fluxC}
\end{center}
\end{figure}

Fig.~\ref{fig:fluxC} shows the comparison of the fluxes for the analytical solution, the LLE slab without bubbles and with a dispersed phase with a=20m$^2$/m$^3$. Results show that the micro phase has no significant effect on the flux, thus the diffusion coefficient is not modified. Moreover, simulations are in well agreement with the analytical solution. In the present case simulations no significant differences arise for different isotopes. Hence, for simplicity, isotopes other that hydrogen have been omitted in Fig.~\ref{fig:fluxC}.

It must be noted that higher specific areas may have an effect on the diffusion process, but the bubbles would be big enough to have an effect on the LLE flow. Therefore, the passive scalar approach would not hold and a two-phase model taking into account gas as a discretized phase should be used, which is out of the scope of the present paper and not the exposed type of system. It is worth to be mentioned that most of the LLE technology applications do not involve large bubbles.


\subsection{Hydrogen transport analysis in a LLE flowing through a sudden expansion in a pipe in the presence of micro bubbles}\label{sec:case4}

In this section, we would like to present a numerical experiment designed to show the impact of vortical structures on hydrogen transport in the presence of micro bubbles. A simplification of geometry from the three-dimensional case, to a 2D axisymmetric case, was used. As it has already been mentioned, the mass transfer model does not affect significantly the amount of absorbed H, which is very low for most of the cases. In a zero-dimensional case, hydrodynamic effects, as well as possible permeation related effects cannot be observed, so the present case is exposed in order to analyse how hydrodynamic structures, like vortices (see, e.g., the hydrodynamic structures in a toroidal oriented manifold of a breeding blanket, exposed in \cite{Morley}), may have a significant impact on H inventory with critical consequences in current nuclear fusion/advanced fission technological designs.

\begin{figure}[ht!]
\begin{center}\vspace{-5pt}
\includegraphics[angle=0,width=1\columnwidth]{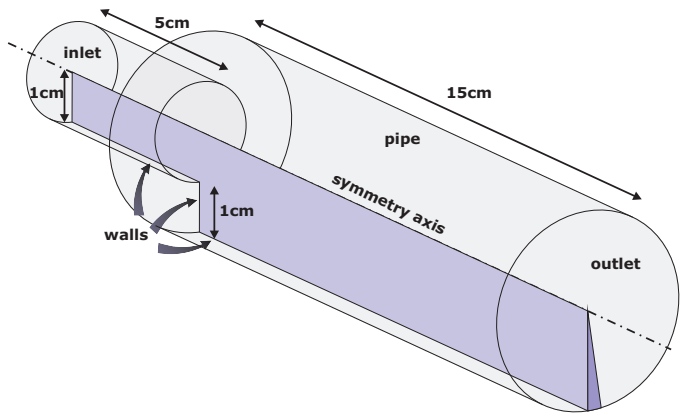}
\caption{2D Axisymmetric BFS simplification of the case geometry and dimensions.}
\vspace{-10pt}
\label{fig:expansion}
\end{center}
\end{figure}

A sudden expansion in a LLE flowing pipe has been chosen in order to show the effect of the well--known vortices or recirculation zones that this type of geometry generates. Many authors, e.g. \citet{Buhler}, who studied the  magneto--hydrodynamic (MHD) effects of LM in a sudden expansion and \citet{Guo} or \citet{Biswas}, have extensively studied and modelled these types of geometries. In addition, sudden expansion backward facing step (BFS) is a widely used benchmark problem for CFD validating purposes, so the vortices in these systems are well known structures. In the present analysis no MHD or thermalhydraulic (TH) effects are considered so as to show only the effect of T absorption into He bubbles. 

Case geometry, meshed in $\sim$10$^{5}$ wedge type cells, and boundaries are shown in Fig.~\ref{fig:expansion}. An expansion ratio of 1.5 has been chosen in order to have a significant recirculation zone. A constant temperature of 723.15~K and 2~bar pressure is set together with the same parameter values as in Sec.~\ref{sec:case1}. Hydrodynamics are simulated with an standard turbulent k--$\omega$ SST OpenFOAM\textsuperscript{\textregistered} solver, suitable for low-Prandtl LMs, at a 0.01m/s inlet constant velocity (Re=200). He and T concentration in the LLE at the inlet is set to a constant value of 0.0004~mol/m$^3$ and 0.001~mol/m$^3$ respectively. He concentration is set to a value slightly over the saturation concentration in order to avoid bubble collapse and to keep bubble growth due to He diffusion very low.

Regarding He bubbles, a constant inlet bubble concentration of 10$^{5}$~bubbles/m$^{3}$ with a mean radius of 4.5$\cdot$10$^{-5}$~m is set, which correspond to a void fraction of 3.5$\cdot$10$^{-5}$. Note that in the present case no nucleation will occur and bubbles enter the system T free.

\begin{figure*}[ht!]
\begin{center}\vspace{-5pt}
\includegraphics[angle=0,width=1\linewidth]{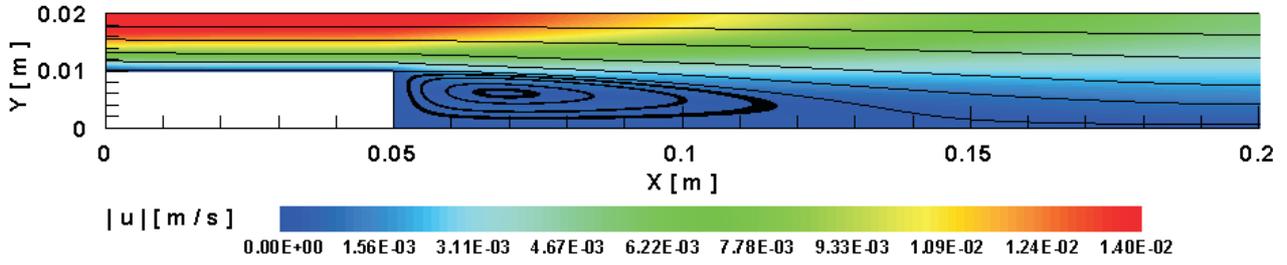}
\caption{BFS stream lines on velocity magnitude field.}
\vspace{-10pt}
\label{fig:U}
\end{center}
\end{figure*}

The hydrodynamic steady state solution is shown in Fig.~\ref{fig:U}. Despite the fact that it is not the aim of this work to analyse the hydrodynamics of the presented system, it should be noted that the simulation shows the formation of the well-known vortices or recirculation zones at the BFS (\citep{Buhler},\citep{Guo} and \citep{Biswas}). Note that this type of sudden expansions are present in fusion reactor LM cooling loops (see, e.g., manifolds in \citep{Morley}).

\begin{figure*}[ht!]
\begin{center}\vspace{-5pt}
\includegraphics[angle=0,width=1\linewidth]{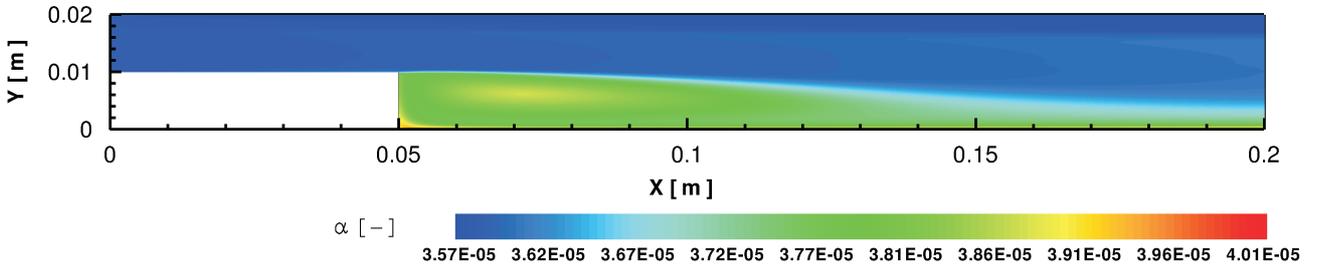}
\caption{BFS void fraction field showing the gas phase accumulation at the vortex.}
\vspace{-10pt}
\label{fig:alpha}
\end{center}
\end{figure*}
\begin{figure*}[ht!]
\begin{center}\vspace{-5pt}
\includegraphics[angle=0,width=1\linewidth]{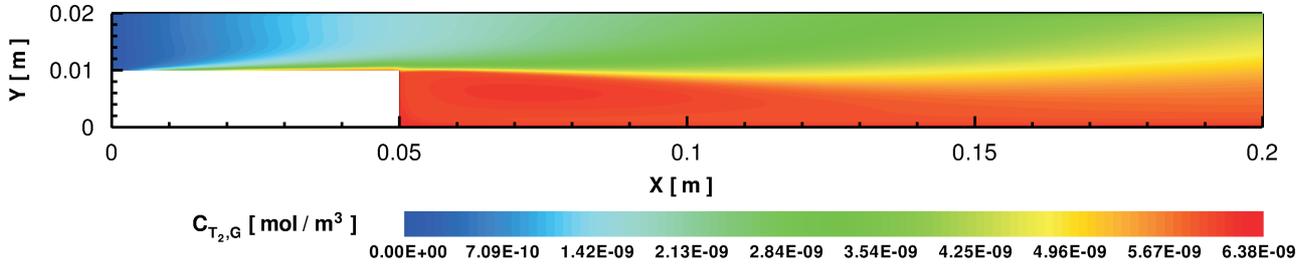}
\caption{BFS molecular T$_2$ concentration inside the bubbles field.}
\vspace{-10pt}
\label{fig:CTG}
\end{center}
\end{figure*}

Once the hydrodynamic steady state solution has been found, He and H transport simulation was run for 6000s as a post-process, but no steady state solution was found as bubbles will keep growing inside the vortex. Hence, such bubbles keep absorbing H, but with a non significant impact on the void fraction as it is shown in Fig.~\ref{fig:alpha} ($<1\%$ due to T and $<10\%$ in volume due to He absorption as $C_{He,LM}>C_{He,LM}^{sat}$). The amount of absorbed H may have a significant impact on the H inventory as shown in Fig.~\ref{fig:CTG} (concentration is expressed per LM volume); the longer the bubbles stay in the vortex, the more T will be removed from the bulk liquid. Moreover, as the vortex is very close to the pipe wall, there may be a substantial effect on the T permeation through the pipe's structural material. Note that H concentration is higher for the bubbles placed at the lower wall. Quantification of such phenomenon is out of the scope of the present paper and it may deserve a dedicated one.

The present case shows from a qualitative and quantitative point of view how vortices may have a significant impact on H inventory in the presence of He bubbles for the presented case. This fact has to be taken into account for design, safety and operation purposes.

\section{Conclusions}

The work presented in this manuscript aims to provide the nuclear technology community with a simple model aiming to shed light on the complex phenomena of hydrogen transport in matter when a gas phase is present, a situation of critical importance in advanced fission systems and current fusion technology designs.
Our model is implementing and taking into account key factors like hydrogen isotopes and helium diffusion and convection in liquid metals, hydrogen absorption into gas bubbles, and helium absorption and bubble growth. To our knowledge, it is the first approach in implementing and adapt to CFD in order to carry out predictive simulations of value in nuclear technology. As far as the authors know, there are no experimental data available. However, \citep{Lee} exposed an experiment that may be used in the future to adjust present model parameters.

Preliminary results obtained by direct application of this model show that, in the presence of a nucleation event or a growing gas phase, like in the case of LLE within a fusion reactor, where Helium and Tritium bubbles are generated due to nuclear reactions from $^6$Li, the hydrogen isotope inventory in the gas phase should be taken into consideration. Hence, some hydrogen extraction processes, like gas-liquid contactors can be assumed to work on the basis that hydrogen isotope concentration in the gas bubbles is the saturation concentration. This may lead to rather low extraction ratios if a large gas-liquid contact area is not used.

Moreover, in the case of a typical geometry for a LLE flowing pipe, we are predicting that the amount of absorbed H may have a significant impact on the H inventory. In this common example, strikingly, H inventory in the presence of He bubbles is highly influenced, being H concentration higher for the bubbles placed at the lower wall vortices. This accumulation may lead to a higher concentration in the structural material and, therefore, a higher permeation rate out of the pipe. Not only a high concentration of hydrogen isotopes may be present in the structural material, leading to material embrittlement, but also hydrogen isotopes leakage may be significant. Thus, our study, although based on a numerical model, suggests a possible accumulation of H in spots present in common system geometries, implying new considerations of design, and possible safety and operation concerns.

%% The Appendices part is started with the command \appendix;
%% appendix sections are then done as normal sections

\appendix

\section{Hydrogen isotope absorption impact on bubble volume} \label{App:App1}
If hydrogen isotopes are absorbed into a He gas bubble, its volume may change. It can be assumed that the amount of absorbed hydrogen isotopes is so small that the volume of the bubbles stays unaffected. However, the effect has been proved to be negligible by implementing a combined equation of state for He and H.

A polynomial fit( Eq.~\ref{eq:HeEoS}) as exposed in \citet{Glasgow}, based on the He EoS by \citet{Trinkaus} has been implemented in the CFD code.
\begin{equation}\label{eq:HeEoS}
Z_{He}=z_{0}^{He}+z_{1}^{He}\rho_{He}^{m}+z_{2}^{He}(\rho_{He}^{m})^{2}+z_{3}^{He}\,(\rho_{He}^{m})^{3},
\end{equation}
where $Z_{He}$ is the He compressibility factor, $\rho_{He}^{m}$ is expressed in mol/dm$^3$ and the $z_{i}^{He}$ factors as a function of temperature $T$ are:
\begin{eqnarray}
\mbox{ \qquad \qquad}\qquad z_{0}^{He} &=&(T/1300)^{0.04},\nonumber\\
\mbox{ \qquad \qquad}\qquad z_{1}^{He} &=& 5.83(1/T)^{0.58},\nonumber\\
\mbox{ \qquad \qquad}\qquad z_{2}^{He} &=& {\log}_{10}(T/800)/(0.69T)^{0.65},\nonumber\\
\mbox{ \qquad \qquad} \qquad z_{3}^{He} &=& 8.6(1/T)^{1.44}.\nonumber\
\end{eqnarray}

A specific H EoS, developed by \citet{Mills}, could be used for pressures over 2~kbar in order to take into account bubble growth due to hydrogen absorption at the very beginning of nucleation. State properties of H$_2$ as a real gas can be expressed by \citet{Mills} EoS in a Berlin-like form.
\begin{equation}
\hat{V}=v_1 P_{b}^{-1/3}+v_2 P_{b}^{-2/3}+v_3 P_{b}^{-1}\label{eq:MillsBerlin}
\end{equation}
where $\hat{V}$ is the molar volume and $P_b$ is the bubble inner pressure. This last equation can be fitted into a Leiden form, which will make easier the combination with a He EoS and its implementation (note that He EoS, eq.~\ref{eq:HeEoS}, is expressed as well in Leiden form).
\begin{equation}
Z_{H_2}=z^{H_2}_{0}+z^{H_2}_{1}\rho_{H_2}^{m}+z^{H_2}_{2}(\rho_{H_2}^{m})^{2}+z^{H_2}_{3}(\rho_{H_2}^{m})^{3}\label{eq:MillsLeiden}
\end{equation}
where $Z_{H_2}$ is the compressibility factor, $\rho_{H_2}^{m}$ is expressed in mol/m$^3$ and the $z_{i}^{H_2}$ are:
\begin{eqnarray}
 z^{H_2}_{0} &=& 6.15449-0.01371T+1,31181\cdotp10^{-5}T^2\nonumber\\
&&-4.39053\cdotp10^{-9}T^3,\nonumber\\
 z^{H_2}_{1} &=& 0.01407-0.27437\cdotp0.9963^T,\nonumber\\
 z^{H_2}_{2} &=& 0.00425-1.36286\cdotp10^{-5}T+2.00322\cdotp10^{-8}T^2\nonumber\\
&&-1.40488\cdotp10^{-11}T^3+3.79781\cdotp10^{-15}T^4,\nonumber\\
z^{H_2}_{3} &=& 2.60566E-7+2.28499\cdotp10^{-9}T-3.84744\cdotp10^{-12}T^2\nonumber\\
&&+2.77354\cdotp10^{-15}T^3-7.48742\cdotp10^{-19}T^4.\nonumber\
\end{eqnarray}

For Berlin and Leiden forms of EoS see \citet{Holborn} and  \citet{Onnes} respectively.

For pressures below 2~kbar the Abel-Noble EoS (eq.~\ref{eq:Abel-Noble}) for H$_2$ could be used. The Abel-Noble EoS constant parameter $b$ for H$_2$ has been determined to be 15.5 with an error of less than 1\% (\citet{Marchi}) from respect to experimental data. Hence, this EoS is perfectly suitable for engineering design purposes.
\begin{equation}
Z_{H_2}=1+\dfrac{Pb}{RT}\label{eq:Abel-Noble}
\end{equation}

For a mixture of H and He the overall compressibility factor $Z_{m}$ can be expressed as follows:
\begin{equation}\label{eq:Mixture}
Z_{m}=y_{H_2,b}\,Z_{H_2}+y_{He,b}\,Z_{He}
\end{equation}
where $y_{H_2,b}$ and $y_{He,b}$ are the molar fractions of H$_2$ and He inside the bubble. 

Combining eq.~\ref{eq:MillsLeiden} and eq.~\ref{eq:Abel-Noble} with eq.~\ref{eq:Mixture}, Young-Laplace mechanical equilibrium equation and eq.~\ref{eq:HeEoS}, an implicit expression for the mixture EoS is found for each H EoS.

The radius of a bubble can be calculated by knowing the He and H concentrations, the LM bulk pressure and the temperature, together with He, H and LM properties.

It must be noted that for H Sievert's constant in LLE, the equilibrium concentration inside the bubbles is always very small when compared to the H concentration in the bulk LM ($C_{H,LM}\sim 10^{6} C_{H_2,G}$). Hence, $C_{H_2,G}$ will never be high enough to have an effect on the bubbles volume. In addition, several simulations have been carried out with and without using the aforementioned combined EoS and it has been found that the impact of hydrogen absorption on the bubble volume is negligible.

\section{Self-Consistent Nucleation Theory implementation} \label{App:App2}
Simple models for He nucleation, bubble growth and transport, were developed and implemented for the \OpF CFD code in \citep{Fradera11}. Implemented models are known to be a good approximation but to extensively overestimate surface tension and nucleation rates. Major improvements to the classical nucleation theory (CNT) homogeneous nucleation (HON) model have been implemented so as to ensure more reliable and predictive simulations.

The Self-Consistent Nucleation Theory (SCT) \citep{Girshick90},  \cite{Girshick91} has gained acceptance due to its good results and simplicity.

The nucleation rate can be expressed as follows:

\begin{equation}
S_{SCT,HON}\ =\ \dfrac{e^\Theta}{\psi}\, S_{CNT,HON}\label{eq:SCTHON}
\end{equation}
where $\psi$ is the supersaturation ratio and $\Theta$ the surface energy of one He atom in the cluster:
\begin{equation}
\Theta\ \equiv\ \dfrac{\sigma s_0}{k_{B}T}\label{eq:SCTcorr}
\end{equation}

and $S_{CNT,HON}$ is the nucleation rate predicted by the CNT and formulated by \citet{Volmer}, \citet{Farkas}, \citet{Becker}, \citet{Zeldovich} and \citet{Frenkel} for homogeneous nucleation.

Moreover, one of the main drawbacks of the aforementioned nucleation theories is the estimation of the surface tension of a cluster or a tiny bubble, which is estimated as that of a planar surface. A simplified relation between the surface tension and the radius of growing bubbles, derived by \citet{Tolman}, have been implemented to take into account the geometry of the bubbles.

\section{Series of resistances flux at the interface} \label{App:App3}

Absorption process can be modelled as a resistance due to diffusion through an stagnant layer around the interface and a resistance at the interface. Assuming both resistances work in series and that the solute cannot accumulate at the interface the fluxes read,

\begin{equation}\label{eq:fluxEq}
J_{H,LM \rightarrow int}=J_{H,int \rightarrow G}
\end{equation}
where $int$ denotes the interface. The diffusion flux can be expressed as follows:
\begin{equation}\label{eq:fluxDiff}
J_{H,LM \rightarrow int}=\dfrac{D_{H,LM}}{r_b}\left(C_{H,int}-C_{H,LM}\right)
\end{equation}
where $r_b$ is teh bubble radius. 
The interface or bubble surface flux reads,
\begin{equation}\label{eq:fluxRec}
J_{H,int \rightarrow G}=k_d p_{H_2} - k_r C_{H,int}^2
\end{equation}

The solution of the quadratic expression that arises from eq~\ref{eq:fluxEq} for $C_{H,int}$ gives the flux expression that takes into account both processes as follows:

\begin{eqnarray}\label{eq:flux2}
&J_{H,LM \rightarrow G}=\dfrac{D_{H,LM}}{\delta}\Bigg[C_{H,LM}-\nonumber\\
&\Bigg(\dfrac{-D_{H,LM}}{2k_r\delta}+\sqrt{\left(\dfrac{D_{H,LM}}{2k_r\delta}\right)^2+\dfrac{-D_{H,LM}C_{H,LM}}{k_r\delta}+\dfrac{k_d p_{H_2}}{k_r}}\Bigg)\Bigg]\nonumber\\
\end{eqnarray}
%

%% References
%%
%% Following citation commands can be used in the body text:
%% Usage of \cite is as follows:
%%   \citep{key}          ==>>  [#]
%%   \cite[chap. 2]{key} ==>>  [#, chap. 2]
%%   \citep{key}         ==>>  Author [#]

%% References with bibTeX database:

\bibliographystyle{model1a-num-names}
\bibliography{references}

%% Authors are advised to submit their bibtex database files. They are
%% requested to list a bibtex style file in the manuscript if they do
%% not want to use model1a-num-names.bst.

%% References without bibTeX database:

% \begin{thebibliography}{00}

%% \bibitem must have the following form:
%%   \bibitem{key}...
%%

% \bibitem{}

% \end{thebibliography}

\end{document}